\documentclass[doublecol]{epl2}
\usepackage{amsmath,amssymb}
\usepackage{graphicx,color,soul}
\usepackage{bm}
\usepackage[colorlinks,linkcolor=magenta,citecolor=magenta]{hyperref}

\newcommand{\dd}{\mathrm{d}}

\newcommand{\pd}[2]{\frac{\partial #1}{\partial #2}}
\newcommand{\mean}[1]{\langle #1 \rangle}
\newcommand{\Int}[1]{\int\dd #1\;}
\newcommand{\IInt}[3]{\int_{#2}^{#3}\dd #1\;}

\renewcommand{\vec}[1]{\mathbf #1}

\newcommand{\al}{\alpha}
\newcommand{\gam}{\gamma}
\newcommand{\eps}{\varepsilon}

\newcommand{\vhi}{\varphi}
\newcommand{\sig}{\sigma}

\newcommand{\x}{\vec r}

\renewcommand{\tx}{\tau_\text{r}}
\newcommand{\nois}{\bm\xi}

\newcommand{\kT}{k_\text{B}T}

\begin{document}

\title{Effective dynamics and fluctuations of a trapped probe moving in a fluid of active hard discs}

\author{Ashreya Jayaram \and Thomas Speck}
\institute{Institute for Theoretical Physics IV, University of Stuttgart, 70569 Stuttgart, Germany}

\abstract{We study the dynamics of a single trapped probe surrounded by self-propelled active particles in two dimensions. In the limit of large size separation, we perform an adiabatic elimination of the small active particles to obtain an effective Markovian dynamics of the large probe, yielding explicit expressions for the mobility and diffusion coefficient. To calculate these expressions, we perform computer simulations employing active Brownian discs and consider two scenarios: non-interacting bath particles and purely repulsive interactions modeling volume exclusion. We keep the probe-to-bath size ratio fixed and vary the propulsion speed of the bath particles. The positional fluctuations of a trapped probe are accessible in experiments, for which we test the prediction from the adiabatic elimination. Although the approximations cause a discrepancy at equilibrium, the overall agreement between predicted and measured probe fluctuations is very good at larger speeds.}

\maketitle

%% ---- introduction ----

\section{Introduction}

The motion of a solvated colloidal particle is due to a myriad of collisions with the surrounding solvent molecules. Observing only the position of the particle necessarily implies a stochastic process, which for a passive solvent is restricted by statistical physics through the Einstein relation (a special case of the more general fluctuation-dissipation theorem~\cite{chandler87}). Due to the scale separation, the collisions become essentially uncorrelated on the time scale of the (much bigger) colloidal particle and typically are modeled as white noise (although deviations can been detected experimentally~\cite{franosch11}). Colloidal probes can also be employed to study local properties of complex fluids and heterogeneous biological materials through the probe's fluctuations~\cite{cicuta07,wilhelm08}. Here one distinguishes between passive and active microrheology, whereby in the later case the probe is driven externally (linearly or oscillatory)~\cite{mizuno07,mizuno08,wilson09}. The response to a driven probe is governed by the deformation of the microstructure, the statistical arrangement of bath particles around the probe, which has been studied in detail for a bath of hard passive particles~\cite{squires05}.

Recently, the concept of \emph{active fluids}, in which constituent particles move autonomously by incessantly transforming residual or stored free energy into directed motion and dissipating heat, has gained traction~\cite{saintillan18,ramaswamy19}. The directed motion can be rotational (``spinners'')~\cite{kokot17,nguyen14} or linear (self-propulsion)~\cite{romanczuk12}, which are distinguished as active chiral~\cite{banerjee17,soni19}, scalar~\cite{buttinoni13}, and polar fluids~\cite{bricard13,chardac21}. The behavior of embedded passive particles in such synthetic active fluids, in addition to bacterial suspensions, has attracted enormous interest~\cite{leptos09,mino13}. For example, asymmetric probe particles embedded in an active fluid can be used to extract useful work~\cite{sokolov10,dileonardo10,speck21}. The statistics of symmetric colloidal probes has been measured in experiments~\cite{wu00,ortlieb19,kanazawa20} and effective equations of motion for the probe particle have been derived~\cite{knezevic20,feng21,tripathi22}. Moreover, the extent to which the fluctuation-dissipation theorem is violated can be exploited to probe the environment~\cite{berthier13,ben-isaac15,maggi17,ye20,shea22}. Brady and coworkers have worked out consequences of a driven probe~\cite{burkholder20,peng22}. Special situations such as (anomalous) tracer diffusion in one dimension~\cite{banerjee22,granek22} and at high densities have been explored theoretically~\cite{reichert21}.

\begin{figure}[t]
  \centering
  \includegraphics{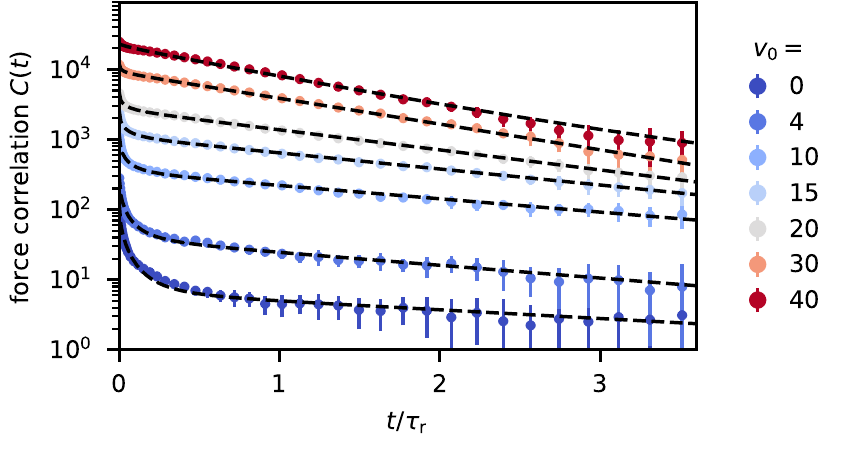}
  \caption{Probe in active bath. (a)~Sketch of the system. A disc-shaped probe is trapped by, e.g., optical tweezers and moving in a bath of active Janus particles. (b)~Force autocorrelations $C(t)$ of a fixed probe for several propulsion speeds $v_0$ of the active bath particles. Lines are fits to Eq.~\eqref{eq:CF} describing a fast relaxation followed by a slower exponential relaxation.}
  \label{fig:force}
\end{figure}

Motivated by recent experiments~\cite{liu20,paul22}, here we study the motion of a trapped passive probe surrounded by self-propelled active particles [Fig.~\ref{fig:force}(a)]. In the experiments, the probe is trapped by optical tweezers and the active particles are strongly confined light-driven Janus colloidal particles moving effectively in two dimensions in a binary near-critical solvent~\cite{gomez-solano17}. Knowing the trap stiffness, forces onto the probe can be related to the displacement from the trap center.

Solon and Horowitz have studied the relationship between diffusion coefficient and mobility of a probe moving in a bath modeled as active Brownian particles~\cite{solon22}. They have compared a passive fluid and an active fluid with constant (relatively small) propulsion speed of the active particles and varied the size ratio. Here we fix the size ratio and investigate a large range of propulsion speeds, determining in independent simulations the force correlations and the effective drag on the probe. We then turn to a trapped particle since this is a relevant experimental strategy to probe complex non-equilibrium environments.

%% ---- model ----

\section{Model}

We study a two-dimensional system composed of $N$ active particles and a single probe trapped in a harmonic potential with stiffness $k$. The overdamped dimensionless equations of motion for the active particles read
\begin{equation}
  \dot\x_k = \eps^{-1}(v_0\vec e_k + \vec F_k) + \sqrt{2\eps^{-1}}\nois_k.
\end{equation}
Here, $\eps\equiv\mu_\text{p}/\mu_\text{a}\ll 1$ is the ratio between the bare mobility $\mu_\text{p}$ of the probe and the bare mobility $\mu_\text{a}$ of the active particles. We measure energies in units of $\kT$, lengths in units of $\sig$ (taken as the diameter of the active particles), and time in units of $\sig^2/(\kT\mu_\text{p})$. The propulsion speed is $v_0$ along the unit directions $\vec e_k$, which undergo rotational diffusion. With $D_\text{r}$ being the (dimensionful) rotational diffusion coefficient, $\tx=\eps\tilde\tx=\kT\mu_\text{p}/(D_\text{r}\sig^2)$ is the dimensionless time over which orientational correlations decay. No-slip boundary conditions (as relevant for self-propelled colloidal particles) imply that rotational and translational diffusion couple with $\tilde\tx=1/3$. Finally, the translational Gaussian noise $\nois_k$ has zero mean and unit variance. In practice, we perform simulations of the active particles with mobility $\eps^{-1}$ and speed $\eps^{-1}v_0$.

The force $\vec F_k=-\nabla_kU$ is composed of the repulsive forces with pair potential $u_\text{b}(r)$ due to neighboring particles and due to the probe with potential $u_\text{p}(r)$. The total potential thus reads $U=\sum_ku_\text{p}(|\x_k-\x|)+\sum_{k<l}u_\text{b}(|\x_k-\x_l|)$. We will study two scenarios: non-interacting active bath particles with $u_\text{b}=0$ and (almost) hard discs for which we employ the repulsive Weeks-Chandler-Andersen potential given by ($i=\text{b,p}$)
\begin{equation}
  \label{eq:wca}
  u_i(r) = 4\epsilon\left[\left(\frac{\sig_i}{r}\right)^{12} - \left(\frac{\sig_i}{r}\right)^6 + \frac{1}{4}\right]\Theta(2^{1/6}\sig_i-r)
\end{equation}
to penalize overlaps between any two particles. Here $\Theta(\cdot)$ is the Heaviside step function and we set $\epsilon=100$ with $\sig_\text{b}=1$ and $\sig_\text{p}=(1+\eps^{-1})/2$, and vary the speed $v_0$. In both scenarios, bath particles and probe interact through their excluded area and the displacement $\x$ of the probe from the trap center evolves according to
\begin{equation}
  \label{eq:eom:p}
  \dot\x = -k\x + \vec F + \sqrt{2}\nois
\end{equation}
with force $\vec F\equiv\sum_k\vec f(\x_k-\x)$ onto the probe and $\vec f=-\nabla u_\text{p}$. Again, the translational Gaussian noise $\nois$ has zero mean and unit variance.

Fluids of interacting repulsive active Brownian discs are known to undergo motility-induced phase separation (MIPS) at sufficiently high speeds, which results in the coexistence of dilute and dense regions~\cite{cates15}. Here we study a moderate global density $\bar\rho=N/A=0.2$ corresponding to a packing fraction of about 16\%. While for speeds $v_0\gtrsim100$ the system crosses the binodal into the two-phase region, for these low packing fractions nucleation of the dense phase is strongly suppressed and the fluid remains homogeneous (albeit metastable)~\cite{richard16}.

%% ---- effective dynamics ----

\section{Effective dynamics}

In the appendix, we project out the bath degrees of freedom to derive
\begin{equation}
  \partial_t\psi = \mu\nabla\cdot(k\x+\mathcal T\nabla)\psi
  \label{eq:psi:fin}
\end{equation}
governing the evolution of the probability distribution $\psi(\x,t)$ to find the probe displaced by $\x$ from the trap center. This equation is isomorphic to a passive probe with mobility $\mu\equiv(1+\zeta)^{-1}$ and effective temperature
\begin{equation}
  \mathcal T \equiv \frac{1+D}{1+\zeta}.
  \label{eq:T:eff}
\end{equation}
While we will see that $\zeta$ is an effective drag due to the bath, the coefficient
\begin{equation}
  D = \IInt{t}{0}{\infty} C(t), \qquad \mean{F_i(t)F_j(0)}_0 = C(t)\delta_{ij}
  \label{eq:D}
\end{equation}
is related to the autocorrelations of the force $\vec F$ onto a \emph{fixed} probe. Due to symmetry, the correlation matrix reduces to a diagonal matrix with function $C(t)$. The adiabatic elimination predicts the fluctuation-dissipation relation $D_\text{eq}=\zeta_\text{eq}$ in thermal equilibrium, which implies $\mathcal T=1$ in our units as expected.

%% ---- force correlations ----

\section{Force autocorrelations}

We first turn to the calculation of $D$ [Eq.~\eqref{eq:D}], for which we consider a fixed probe ($\dot\x=0$) immersed in the active bath. We perform Brownian dynamics simulations of the active particles moving in a square box of edge length $L$ with periodic boundary conditions. Throughout, we set the mobility ratio to $\eps=0.1$ corresponding to a size ratio of $10$.

Figure~\ref{fig:force}(b) reveals that the numerical force correlations $C(t)$ exhibit a two-step decay, which is well fitted by the function
\begin{equation}
  C(t) = \frac{D_\text{s}}{\tau_\text{s}}e^{-t/\tau_\text{s}} + \frac{D_\text{f}}{2\tau_\text{f}}e^{-\sqrt{t/\tau_\text{f}}}.
  \label{eq:CF}
\end{equation}
This functional form can be interpreted as a slow exponential relaxation on timescale $\tau_\text{s}$ separated from a spectrum of fast exponential timescales giving rise to a stretched exponential with exponent $\approx1/2$. While the slow timescale should be related to diffusion, the origin of the fast timescale is the interactions between probe and bath particles since the non-interacting bath particles show the same two-step decay. Note that our data is not compatible with a power-law decay as suggested in~\cite{solon22}. For large speeds, the fast relaxation effectively vanishes. For even higher speeds $v_0\gtrsim100$, the statistics become insufficient even for intermediate times with a narrow window of fast decay.

\begin{figure}[t]
  \centering
  \includegraphics{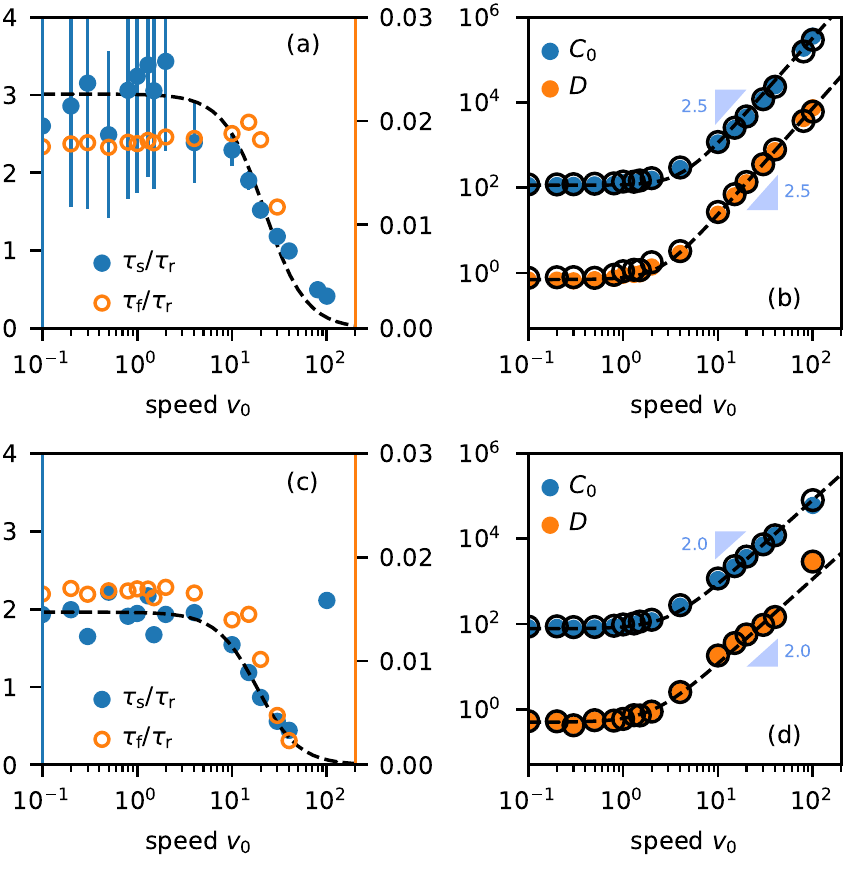}
  \caption{Analysis of force correlations. (a)~Fitted relaxation times $\tau_\text{s}$ (left axis) and $\tau_\text{f}$ (right axis) divided by the rotational relaxation time $\tx$ as a function of speed $v_0$. While $\tau_\text{s}$ is of order $\tx$, the fast relaxation $\tau_\text{f}\ll\tau_\text{s}$ is two orders of magnitude faster. The dashed line is a fit to Eq.~\eqref{eq:tau:fit}. (b)~The increase of the static force correlations $C_0$ and the integral $D$ follow a power law as a function of propulsion speed with exponent $\nu=2.5$ (dashed lines). The open circles indicate the values calculated from the fits [Eq.~\eqref{eq:CF}]. (c,d)~Same analysis but now for a bath of non-interacting active discs.}
  \label{fig:fit}
\end{figure}

In Fig.~\ref{fig:fit}(a), we plot the fitted relaxation times. While $\tau_\text{s}$ is of order $\tx$ and thus determined by the decay of the orientational correlations of the active particles around the probe, the fast relaxation times $\tau_\text{f}$ are two orders of magnitude smaller. Both time scales have vanishing slope for small speeds (not withstanding the large uncertainty of $\tau_\text{f}$) and decay for large propulsion speed, which we will see corresponds to an ``active thinning'' of the bath. Assuming that the slow decorrelation of forces is determined by an effective diffusion of active particles leads to
\begin{equation}
  \tau_\text{s}(v_0) = \frac{\ell^2}{D_\text{eff}(v_0)} = \frac{\tau_\text{eq}}{1+(v_0/v_\tau)^2}
  \label{eq:tau:fit}
\end{equation}
with length $\ell\equiv\sqrt{D_\text{eff}(0)\tau_\text{eq}}$ and assuming that the diffusion coefficient increases quadratically. This functional form yields a reasonable fit to the data for interacting active discs [Fig.~\ref{fig:fit}(a)] with $\tau_\text{eq}\simeq3.0$ and crossover speed $v_\tau\simeq 22.2$. A bath of non-interacting active discs ($u_\text{b}=0$) behaves qualitatively similar as shown in Fig.~\ref{fig:fit}(c) with slight different fit parameters $\tau_\text{eq}\simeq2.0$ and $v_\tau\simeq18.5$.

In Fig.~\ref{fig:fit}(b), we show the static force correlations $C_0$ together with the time integral of the force correlations $D=D_\text{s}+D_\text{f}$ as a function of propulsion speed $v_0$. Interestingly, both quantities seem to increase as a power law (i.e., $D-D_\text{eq}\propto v_0^\nu$) with an exponent close to $\nu\simeq5/2$. We also calculate the corresponding values from the fitted correlation functions [Eq.~\eqref{eq:CF}], which agree well with the directly determined values. For non-interacting active discs [Fig.~\ref{fig:fit}(d)], the most notable difference is the change of exponent for the increase of the static and integrated force correlation $C_0$ and $D$, respectively, which are now better described by an exponent $\nu\simeq 2$.

%% ---- microrheology ----

\section{Driven probe}

In order to calculate the mobility, we now consider a probe moving according to
\begin{equation}
  \vec u = \dot\x = f_\text{p}\vec e_x+\vec F
  \label{eq:fp}
\end{equation}
driven by a constant force $f_\text{p}$ along the direction $\vec e_x$. Introducing the mobility $\mu$ through $\mean{\vec u}=\mu f_\text{p}\vec e_x$ and writing $\mean{\vec F}=-\zeta\mean{\vec u}$, we confirm the relation $\mu=(1+\zeta)^{-1}$ between mobility $\mu$ and the drag coefficient $\zeta$.

\begin{figure}[t]
  \includegraphics{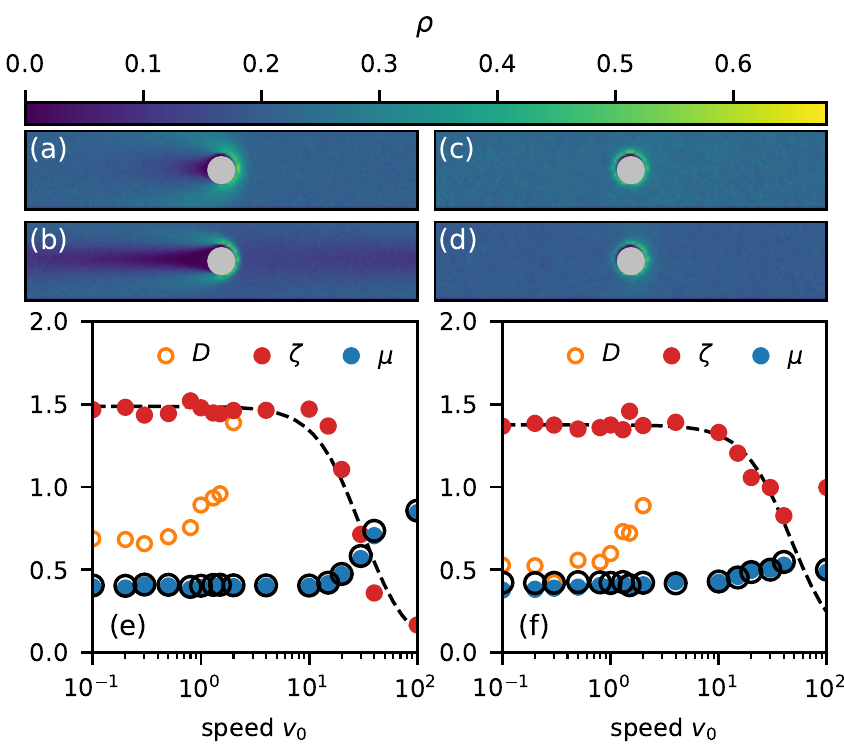}
  \caption{Driven probe. Density distributions of active particles around the probe for (a)~$v_0=10$, $f_\text{p}=40$, (b)~$v_0=10$, $f_\text{p}=200$, (c)~$v_0=200$, $f_\text{p}=40$, and (d)~$v_0=200$, $f_\text{p}=200$. (e)~Drag coefficient $\zeta$ and mobility $\mu$ as a function of speed for interacting bath particles. The dashed line is a fit to Eq.~\eqref{eq:zeta:fit}. The open circles indicate $(1+\zeta)^{-1}$, which agree with the mobility as predicted. Also shown are the integrated force correlations $D$. (f)~Same as panel (e) but for non-interacting bath particles.}
  \label{fig:rheo}
\end{figure}

To accommodate the anticipated deformation of the average arrangement of bath particles around the probe, simulations are performed in an elongated box with dimensions $L_x=5L_y$. For small $v_0$, as shown in Fig.~\ref{fig:rheo}(a,b), the dragged probe witnesses an accumulation of active particles ahead of it and leaves behind a pronounced wake similar to a bath of passive hard discs~\cite{squires05}. The span of the wake increases as $f_\text{p}$ increases. For larger $v_0$, the distribution of particles around the probe remains homogeneous for small $f_\text{p}$ and becomes inhomogeneous for larger driving force $f_\text{p}$, cf. Fig.~\ref{fig:rheo}(c,d).

We perform a series of simulations for each propulsion speed $v_0$ increasing the driving force $f_\text{p}$. We determine
\begin{equation}
  \mu = \lim_{f_\text{p}\to0}\frac{\mean{u_x}}{f_\text{p}}, \qquad
  \zeta = -\lim_{f_\text{p}\to0}\frac{\mean{F_x}}{\mean{u_x}}
\end{equation}
from independent measurements of speed and force. These are plotted in Fig.~\ref{fig:rheo}(e) for interacting and in Fig.~\ref{fig:rheo}(f) for non-interacting bath particles. The relation $\mu=(1+\zeta)^{-1}$ is obeyed as expected. For $v_0\to0$, we expect $D_\text{eq}=\zeta_\text{eq}$, which, however, is not obeyed. This failure indicates that the approximations underlying Eq.~\eqref{eq:psi:fin} are not (yet) fulfilled for the mobility ratio $\eps=0.1$. Interestingly, this holds for both interacting and non-interacting bath particles.

The drag coefficient $\zeta$ remains approximately constant for small $v_0$ and drops beyond $v_0\gtrsim 10$. Active baths at high propulsion speed thus exert a reduced drag (``active thinning'') in agreement with previous results obtained by Burkholder and Brady~\cite{burkholder20}. We observe that the data is well fitted by
\begin{equation}
  \zeta(v_0) = \frac{\zeta_\text{eq}}{1+(v_0/v_\zeta)^2}
  \label{eq:zeta:fit}
\end{equation}
in analogy with the reduction of the slow timescale [Eq.~\eqref{eq:tau:fit}]. Again, both interacting and non-interacting bath particles behave similarly.

%% ---- MSD ----

\section{Steady state fluctuations}

Having calculated the two coefficients $D$ and $\zeta$, we now turn to the steady-state fluctuations $\chi\equiv\mean{\x^2}$ of the moving probe confined by a quadratic potential. In equilibrium, the partition function
\begin{equation}
  Z(k) = \Int{^2\x\dd^2\x_1\cdots} e^{-(U+\frac{1}{2}k\x^2)} = \frac{2\pi}{k}Z_\text{b}
\end{equation}
factorizes after shifting positions $\x_k\to\x_k+\x$. The second moment $\chi_\text{eq}=\mean{\x^2}_\text{eq}=-2\partial_k\ln Z(k)=2/k$ is thus independent of the bath--probe interactions. Multiplying Eq.~\eqref{eq:psi:fin} by $\x^2$ and performing integration by parts twice, we obtain $\chi=(2/k)\mathcal T$, which reduces to the equilibrium result for $v_0=0$. Note that from Eq.~\eqref{eq:eom:p} we can derive the exact relation $k\mean{\x^2}=2+\mean{\x\cdot\vec F}$, which implies another expression $\mathcal T=1+\mean{\x\cdot\vec F}/2$ for the effective temperature.

\begin{figure}[t]
  \centering
  \includegraphics{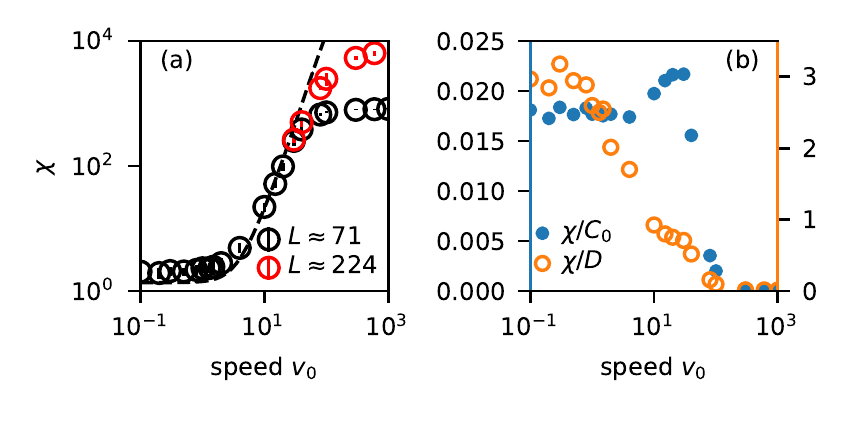}
  \caption{Probe fluctuations in a trap with stiffness $k=1$. (a)~Averaged squared distance $\chi=\mean{\x^2}$ to the trap center as a function of propulsion speed $v_0$. Shown are two system sizes. The dashed line is the prediction $2\mathcal T$ from the effective dynamics together with Eq.~\eqref{eq:T:eff}. (b)~Reduced fluctuations with respect to $C_0$ and $D$.}
  \label{fig:msd}
\end{figure}

In Fig.~\ref{fig:msd}(a), we plot $\chi$ as a function of propulsion speed for $k=1$. Simulations are again performed in a square box but now the probe moves according to Eq.~\eqref{eq:eom:p}. The positional fluctuations start at $\chi_\text{eq}=2$ and then rise quickly before saturating at large speeds. This saturation is a finite-size effect as the fluctuations $\chi\sim L^2$ reach the box size, which is confirmed by simulating a larger system size shifting the saturation to larger values of $\chi$ [red symbols in Fig.~\ref{fig:msd}(a)]. We also plot the prediction $2\mathcal T$ from the effective dynamics using the functional forms for $D$ and $\zeta$ in Eq.~\eqref{eq:T:eff} together with the fitted parameters. This prediction agrees very well with the simulation results except for small speeds, where it underestimates the correct value of $2$ due to the violation of the fluctuation-dissipation relation $D_\text{eq}=\zeta_\text{eq}$.

In Fig.~\ref{fig:msd}(b), we show how the fluctuations $\chi$ behave as a function of the static and integrated force correlation. We find a constant factor $\chi/C_0\simeq0.018$ up to speeds $v_0\lesssim 10$, which shows that the static force correlations are well correlated with the positional fluctuations. This correlation is much less pronounced with the integrated force correlations $D$.

%% ---- conclusions ----

\section{Conclusions}

Probing complex environments out of equilibrium, such as living matter, requires reliable models to interpret the data. Observing an optically trapped probe is a feasible experimental strategy, with the probe fluctuations encoding the activity of the environment. Through an adiabatic elimination of bath particles, we have derived the effective memoryless Markovian dynamics [Eq.~\eqref{eq:psi:fin}] of a probe particle moving in an external potential (here a quadratic potential) in a bath of active particles. The small expansion parameter is the mobility ratio and the result should hold for any propulsion speed of the active bath particles. To assess the validity of the effective dynamics, we have performed Brownian dynamics simulations of hard discs modeled through the WCA potential Eq.~\eqref{eq:wca}. Interestingly, we find only rather small differences between non-interacting and interacting bath particles. Our system is characterized by a linear response regime up to propulsion speeds $v_0\simeq10$, beyond which it undergoes ``active thinning'' characterized by a reduced drag coefficient and a power-law increase of force correlations and positional fluctuations. Curiously, the exponent for interacting bath particles seems to be closer to $\nu=5/2$ while for non-interacting it is $\nu=2$ more in line with the expected increase of the effective diffusion $D_\text{eff}$ of the active bath particles. It would be interesting to see if discontinuous thinning is possible as has been observed for strong probe-bath attractions~\cite{wulfert16}.

The effective dynamics is governed by the force correlations, which display a two-step decay to a slow exponential decay [Fig.~\ref{fig:force}(b)]. In equilibrium, the integrated force correlations should equal the drag coefficient, which is not obeyed in the data for this system [Fig.~\ref{fig:rheo}(e,f)], presumably due to replacing the exact expression [Eq.~\eqref{eq:D:theo}] by an integral over the force correlations [Eq.~\eqref{eq:D:approx}]. Another possibility is that non-Markovian effects are not negligible, leading to a generalized Langevin equation for which we would have to determine the memory kernel~\cite{hijon10,jung17,meyer20,ayaz22}.

%% ---- acknowledgments ----

\begin{acknowledgments}
  We acknowledge funding by the Deutsche Forschungsgemeinschaft (DFG) within collaborative research center TRR 146 (Grant No. 404840447). Computations have been performed on the supercomputer MOGON II (ZDV Mainz).
\end{acknowledgments}

% ---- appendix ----

\section{Appendix}

For completeness, we now derive the effective evolution equation~\eqref{eq:psi:fin} for the probability $\psi(\x,t)$ of the probe. We perform an adiabatic elimination~\cite{vankampen86,speck04} similar to Ref.~\cite{solon22} but for overdamped probe dynamics and fully accounting for the small expansion parameter $\eps$.

\subsection{Projection formalism}

It will be helpful to explicitly introduce the probe velocity $\vec u$ as an auxiliary variable so that the probe equations of motion read $\dot\x=\vec u$ and
\begin{equation}
  \gam^{-1}\dot{\vec u} + \vec u = -k\x + \vec F + \sqrt{2}\nois
\end{equation}
with relaxation time $\gam^{-1}$. We recover the overdamped equation~\eqref{eq:eom:p} in the limit $\gam^{-1}\to0$. We measure the positions $\x_k=\x+\x_k'$ of bath particles with respect to the probe. The joint probability $\Psi(\{\x_k',\vhi_k\},\x,\vec u;t)$ evolves according to the Smoluchowski equation
\begin{equation}
  \partial_t\Psi = \mathcal L_\text{p}\Psi + \eps^{-1}\mathcal L_0\Psi
\end{equation}
with differential operators
\begin{gather}
  \label{eq:Lp}
  \mathcal L_\text{p} = -\vec u\cdot\nabla + \mathcal L_u + \gam(k\x-\vec F)\cdot\pd{}{\vec u} + \vec u\cdot\sum_k\nabla_k', \\
  \label{eq:La}
  \mathcal L_0 = \sum_k \left[ -\nabla_k'\cdot(v_0\vec e_k+\vec F_k-\nabla_k') + \frac{1}{\tilde\tx}\pd{^2}{\vhi_k^2} \right],
\end{gather}
and $\mathcal L_u=\partial_{\vec u}\cdot(\gam\vec u+\gam^2\partial_{\vec u})$. The advantage of introducing the speed $\vec u$ is that $\vec F_k$, and thus $\mathcal L_0$, are independent of the explicit probe position $\x$ which simplifies some of the following calculations at the expense of having to eliminate the probe speed afterwards.

Exploiting that $\eps$ is small, we aim to eliminate the bath particles and derive an effective evolution equation for the probe. The bath particles assume a stationary distribution $\phi_0(\{\x_k',\vhi_k\})$ defined through $\mathcal L_0\phi_0=0$. Note that $\mathcal L_0$ only acts on relative positions and orientations of the bath particles. Our main tool will be the projection operator
\begin{equation}
  \mathcal P \circ = \phi_0\prod_k\Int{^2\x_k'}\IInt{\vhi_k}{0}{2\pi} \circ
\end{equation}
onto the stationary distribution $\phi_0$ so that $\mathcal P\phi_0=\phi_0$. Clearly, $\mathcal L_0\mathcal P=\mathcal P\mathcal L_0=0$ and we can split
\begin{gather}
  \label{eq:eom:psi0}
  \partial_t\Psi_0 = \mathcal P\mathcal L_\text{p}(\Psi_0+\Psi_1) \\
  \partial_t\Psi_1 = (1-\mathcal P)\mathcal L_\text{p}\Psi_0 + [(1-\mathcal P)\mathcal L_\text{p}+\eps^{-1}\mathcal L_0]\Psi_1
\end{gather}
with $\Psi_0=\mathcal P\Psi$ and the remainder
\begin{equation}
  \label{eq:psi1}
  \Psi_1 = \Psi - \Psi_0 = -\eps\mathcal L_0^{-1}(1-\mathcal P)\mathcal L_\text{p}\Psi_0 + \mathcal O(\eps^2)
\end{equation}
expanded to lowest order of $\eps$. Plugging this expression back into Eq.~\eqref{eq:eom:psi0} leads to an evolution equation for $\Psi_0$ alone.

\subsection{Eliminating the bath particles}

We start by writing
\begin{equation}
  \Psi_0 = \mathcal P\Psi = \phi_0(\{\x_k',\vhi_k\})\psi^u(\x,\vec u,t),
\end{equation}
where $\psi^u$ is the marginal probability of probe position and speed that we are interested in. We now inspect all terms to find an evolution equation for $\psi^u$ alone. Turning to $\mathcal P\mathcal L_\text{p}\Psi_0$, the first two terms factorize and $\mathcal P\nabla_k'=0$, leaving
\begin{equation}
  \phi_0\pd{}{\vec u}\cdot\gam(k\x-\mean{\vec F}_0)\psi^u
\end{equation}
with the average force $\mean{\vec F}_0$ onto the probe. Since the bath particles see a stationary probe they assume a symmetric distribution and this force vanishes, $\mean{\vec F}_0=0$. We thus find
\begin{equation}
  \label{eq:P:psi0}
  \mathcal P\mathcal L_\text{p}\Psi_0 = \phi_0 \left[-\vec u\cdot\nabla + \mathcal L_u + \gam k\x\cdot\pd{}{\vec u}\right]\psi^u
\end{equation}
and
\begin{equation}
  \tilde\Psi = (1-\mathcal P)\mathcal L_\text{p}\Psi_0 = -\phi_0\gam\vec F\cdot\pd{\psi^u}{\vec u} + \phi_0\vec G\cdot\vec u\psi^u
\end{equation}
with $\vec G(\{\x_k',\vhi_k\})\equiv\sum_k\nabla_k'\ln\phi_0$.

The next step is to consider
\begin{equation}
  \mathcal P\mathcal L_\text{p}\Psi_1 = -\eps\mathcal P\mathcal L_\text{p}\mathcal L_0^{-1}\tilde\Psi
\end{equation}
inserting Eq.~\eqref{eq:psi1}. All terms of $\mathcal L_\text{p}$ [Eq.~\eqref{eq:Lp}] that do not involve $\x'$ commute with the projector and drop out since $\mathcal P\Psi_1=0$ has to vanish, as does the last term. The remaining terms are
\begin{equation}
  \mathcal P\mathcal L_\text{p}\Psi_1 = \eps\phi_0\sum_{ij}\left[\gam\zeta_{ij}\pd{}{u_i}u_j\psi^u + \gam^2 D_{ij}\pd{^2\psi^u}{u_i\partial u_j}\right]
\end{equation}
with tensors
\begin{equation}
  D_{ij} \equiv -\mean{F_i\mathcal L_0^{-1}F_j}_0, \qquad
  \zeta_{ij} \equiv \mean{F_i\mathcal L_0^{-1}G_j}_0
  \label{eq:D:theo}
\end{equation}
contributing to the drift and diffusion terms, respectively. Since our original setup is rotationally symmetric, we expect that both reduce to diagonal tensors $D_{ij}=D\delta_{ij}$ and $\zeta_{ij}=\zeta\delta_{ij}$.

Practically, the expressions Eq.~\eqref{eq:D:theo} involving the inverse evolution operator of the bath are not very useful. To convert these into integrals of correlation functions, we insert unity, $1=\IInt{s}{0}{\infty}\delta(s)$, and use that $e^{\mathcal L_0s}\to-\mathcal L_0^{-1}\delta(s)$ in the limit of perfect scale separation when the correlation function approaches a $\delta$-distribution. With this approximation, we transform
\begin{equation}
  D_{ij} = -\mean{F_i\mathcal L_0^{-1}F_j}_0 \approx \IInt{s}{0}{\infty}\mean{F_ie^{\mathcal L_0s}F_j}_0
  \label{eq:D:approx}
\end{equation}
into the integral of the force correlations of a fixed probe that can be measured in simulations. To make contact with the simulations, we define $\eps D\to D$ and $\eps\zeta\to\zeta$ involving $(\eps^{-1}\mathcal L_0)^{-1}$ so that the bath corresponds to active discs with speed $\eps^{-1}v_0$ and (dimensionless) mobility $\eps^{-1}$.

For passive systems, we have $\phi_0\propto e^{-U(\{\x_k'\})}$ up to a normalization constant. Calculating $\vec G$, the sum of the interparticle forces between bath particles vanishes (Newton's 3rd law) and
\begin{equation}
  \vec G_\text{eq} = \sum_{k=1}^N \vec F_k = -\vec F
\end{equation}
is given by the force on the probe. For passive systems we thus have $\zeta_\text{eq}=D_\text{eq}$ in agreement with the fluctuation-dissipation theorem.

Putting everything together, we find the evolution equation
\begin{equation}
  \partial_t\psi^u = (\mathcal L_u' + \mathcal L_\text{p}')\psi^u
\end{equation}
for the marginal distribution $\psi^u(\x,\vec u,t)$ with (intermediate) operators
\begin{gather}
  \mathcal L_\text{p}' = -\vec u\cdot\nabla + \gam k\x\cdot\pd{}{\vec u}, \\
  \mathcal L_u' = \pd{}{\vec u}\cdot\left[\gam\left(1+\zeta\right)\vec u+\gam^2\left(1+D\right)\pd{}{\vec u}\right].
\end{gather}

\subsection{Eliminating probe speed}

The final step is to eliminate the speed $\vec u$ in the limit $\gam^{-1}\to0$. This step is performed in analogy to the elimination of bath particles through factorizing $\psi^u(\x,\vec u,t)=f_0(\vec u)\psi(\x,t)$ with $\mathcal L_u'f_0=0$. It is straightforward to show that the solution is
\begin{equation}
  f_0(\vec u) = \frac{2\pi}{\al}e^{-\al\vec u^2/2}, \qquad \pd{f_0}{\vec u} = -\al\vec uf_0
\end{equation}
with coefficient
\begin{equation}
  \al \equiv \gam^{-1}\frac{1+\zeta}{1+D}.
\end{equation}
The orthogonal component reads [cf.~Eq.~\eqref{eq:psi1}]
\begin{equation}
  \psi_1 \approx -\mathcal L_u'^{-1}(1-\mathcal P_u)\mathcal L_\text{p}' f_0\psi
\end{equation}
with projection operator $\mathcal P_u$ onto $f_0$. Let us look at $\mathcal L_\text{p}'f_0=-f_0\vec u\cdot(\nabla+\gam\al k\x)$ and thus $\mathcal P_u\mathcal L_\text{p}'f_0=0$. Next, $\mathcal L_u'\vec uf_0=-\gam(1+\zeta)\vec uf_0$
implies
\begin{equation}
  \mathcal L_u'^{-1}\vec uf_0 = -\gam^{-1}(1+\zeta)^{-1}\vec uf_0
\end{equation}
leading to
\begin{multline}
  \psi_1 = (\mathcal L_u'^{-1}\vec uf_0)\cdot(\nabla+\gam\al k\x)\psi \\
  = -\gam^{-1}(1+\zeta)^{-1}\vec uf_0\cdot(\nabla+\gam\al k\x)\psi.
\end{multline}
Whence
\begin{multline}
  \mathcal P_u\mathcal L_\text{p}'\psi_1 = -\gam^{-1}(1+\zeta)^{-1}f_0 \times \\ \Int{^2\vec u} f_0(-\vec u\vec u\cdot\nabla+\gam k\x-\gam\al k\x\cdot\vec u\vec u)\cdot(\nabla+\gam\al k\x)\psi \\
  = f_0(1+\zeta)^{-1}\nabla\cdot[(\gam\al)^{-1}\nabla+k\x]\psi
\end{multline}
employing the Gaussian integral $\Int{^2\vec u}u_iu_jf_0=\al^{-1}\delta_{ij}$. We can now read off the final evolution equation~\eqref{eq:psi:fin}, which is independent of $\gam$.

%% ---- bibliography ----

\end{document}